\documentstyle[aps,pre,preprint]{revtex}

\tightenlines
\begin{document}
\author{M.S. Hussein and M.P. Pato}
\address{Nuclear Theory and Elementary Particle Phenomenology Group\\
Departamento de F\'{\i}sica Nuclear, Instituto de F\'{\i}sica, Universidade de 
S\~{a}o Paulo\\
CP 66318, 05315-970 S\~{a}o Paulo, SP, Brazil}
\title{Fractal Structure of Random Matrices\footnote{Supported in part by the 
CNPq - Brazil and FAPESP}}
\maketitle

\begin{abstract}
A multifractal analysis is performed on the universality classes of random
matrices and the transition ones. Our results indicate that the eigenvector
probabi;ity distribution is a linear sum of two $\chi ^{2}$-distribution
throughout the transition between the universality ensembles of random
matrix theory and Poisson.
\end{abstract}

\newpage

The Anderson localization\cite{Anderson} is a wave phenomenon characterized
by a destructive interference that gives rise to unaccessible regions in the
configuration space of a physical system. It occurs in many situations\cite
{Kinnon} and, in particular, in condensed matter physics, it is the
responsible for the metal-insulator phase transition (MIT) caused by an
increasing of disorder in a quantum disordered system, in which, as a
consequence, the material undergoes a transformation from a metallic to an
insulator phase. The eigenstates of the system, in the metallic phase,
extend over all the available space and, in the other hand, at the insulator
side, it is localized around the impurities. This situation implies, as the
transformation proceeds, in a modification of the fractal dimension of the
wavefunctions. This aspect of the transition has been studied throw the use
of the multifractal analysis\cite{Peliti,Janssen} which was introduced some
years ago\cite{Kada}.

Random matrix ensembles are another powerful theoretical tool to study
transitions from extended to localized states\cite{Guhr}. The phase of the
extended states, i.e., the metallic phase in the MIT case, is approached by
the universal ensembles of random matrix theory (RMT)\cite{Meht}, namely,
the Gaussian Orthogonal Ensemble (GOE), if the system has time-reversal
invariance, and the Gaussian Unitary Ensemble (GUE), if not. On the other
hand, the insulator phase, where the states become localized, can be
simulated by a Poissonian ensemble. Accordingly, the energy levels of the
strong mixing metallic states are expected to follow Wigner-Dyson statistics
of RMT, while the levels of the uncorrelated localized states of the
insulator phase, have fluctuations that follow the Poisson statistics. The
Maximum Entropy Principle (MEP) has been used to generate matrix ensembles 
\cite{Pato1} that make the connection between two universal limiting
situations, e.g., RMT and Poisson. The joint probability distribution of the
matrix elements of the interpolating ensembles obtained are given by\cite
{Pato2}

\begin{equation}
P(H,\beta ,\alpha )=K_{N}\exp [-\alpha _{0}Tr(H^{2})]\exp \{-\beta
Tr[(H-H_{0})^{2}]\}  \label{1}
\end{equation}
where $N$ is the dimension of the matrix, $K_{N}$ is the normalization
constant, $\alpha _{0}$ is fixed by a choice of units and $\beta $ is the
parameter that controls the transition. When $\beta $ varies from zero to
infinity, the ensemble undergoes a transition from the RMT ensemble to the
ensemble of $H_{0}$. By choosing $H_{0}$ to be a diagonal, we have the
desired transition from RMT to Poisson. This connection between RMT and
Poisson is not expected to be universal and, in fact, other possible
interpolating ensembles have already been proposed, e.g., band matrices\cite
{Casati} and U(N) invariant ensembles\cite{Muttalib}. However, the above
formalism has the advantage that the chaoticity parameter $\beta $ is easily
expressed in terms of the coupling constant\cite{Pato1,Pato3}. In this
letter, we extend the multifractal analysis to the abstract space of random
matrices to investigate statistical properties of the eigenstates of this
interpolating ensemble.

Multifractal is a mathematical object that is not characterized by a unique
dimension but by an infinite spectrum of dimensions\cite{Vulpi}. For
example, in a classical dynamical system, the probability given by the
frequency with which different ''cells '' of a partition of the strange
atractor is visited, in the time evolution of a chaotic system, is a
multifractal. In the quantum case, chaoticity may be defined as the
situation in which the eigenfunctions spread uniformly over all components
with respect to any basis. Of course, this reflects the fact that the system
does not have any other good quantum numbers but the energy and it is the
quantum mechanical equivalent of the absence of integrals of motion in the
classical case. This property may be considered as an statement about the
dimension of the wave function. On the other hand, when transitions towards
regularity occur, one should expect that the presence of conserved
quantities implies in some localization that should be followed by
modifications in the respective dimensions of the eigenstates. This can be
seen in Fig. 1, where it is plotted the logarithm of the components of a
random eigenstate of the above ensemble as a function of its label,
calculated at a critical value of the parameter $\beta $\cite{Pato2}. It is
also shown in the figure, a Gaussian fit that makes clear the localization.
It can be seen that only a small fraction of the components contribute to
the normalization. In this sense the state does not occupy all the available
space. Here the support is given by the states of the basis and the
probability distribution by the way these basis states are populated by the
eigenstates of the Hamiltonian.

The critical behavior of the above transition ensemble has been defined in
Ref. \cite{Pato2}applying Shannon's concept of entropy to the eigenstates
components treated as probabilities. As a start point to introduce the
multifractal formalism, we observe that this entropy is a particular case of
Tsallis generalized entropy\cite{Tsallis} of a probability distribution $%
p_{i},$ with $i=1,N_{p},$ which in terms of the partition function

\begin{equation}
\chi (q)=\sum\limits_{i=1}^{N_{p}}p_{i}^{q}  \label{4}
\end{equation}
is defined by 
\begin{equation}
S_{q}=\frac{1-\chi (q)}{q-1}.  \label{6}
\end{equation}
for any real $q.$ Associated with $S_{q}$, a spectrum of dimension
functions, $D_{q},$ can then be introduced as\cite{Kada}

\begin{equation}
D_{q}=\lim\limits_{l\rightarrow 0}\frac{\ln \left[ 1-\left( q-1\right) S_{q}%
\right] }{(q-1)\ln l},  \label{9}
\end{equation}
where $l$ is a characteristic size associated with the partition.

Some positive integer values of $q$ have an immediate interpretation. Thus, $%
D_{0}=-\frac{\ln N_{c}}{\ln l},$ where $N_{c}$ is the number of occupied
cells, i.e., those with probability different from zero, gives the fractal
dimension of the support. For $D_{1}$ we obtain

\begin{equation}
D_1=\frac{S_1}{\ln l}  \label{12}
\end{equation}
where

\begin{equation}
S_{1}=-\sum\limits_{i}p_{i}\ln p_{i}  \label{14}
\end{equation}
is, by definition the Shannon's information entropy, and $D_{1}$ is the
information dimension. $D_{2}$ is the correlation dimension.

We now assume the scaling $p_{i}\sim l^{\alpha ^{\prime }}$ and, also, that
the exponents $\alpha ^{\prime }$ vary continuously, as $i$ is varied$,$
with a certain distribution $\rho \left( \alpha ^{\prime }\right) $ that
scales with $l^{f\left( \alpha ^{\prime }\right) }.$ Thus the partition
function can be transformed into an integral as

\begin{equation}
\chi (q)=\int d\alpha ^{\prime }\rho \left( \alpha ^{\prime }\right)
l^{q\alpha ^{\prime }-f\left( \alpha ^{\prime }\right) }  \label{16}
\end{equation}
Since $l$ is a small quantity, this integral will be determined
asymptotically by the value $\alpha ^{\prime }=\alpha $ that minimizes the
exponent of $l$ in the integrand. This yields

\begin{equation}
D_q=\frac 1{q-1}\left[ q\alpha -f\left( \alpha \right) \right]  \label{18}
\end{equation}
from which we can deduce

\begin{equation}
\alpha =\frac{d}{dq}\left[ \left( q-1\right) D_{q}\right]  \label{20}
\end{equation}
These two equations give $\alpha $ and $f\left( \alpha \right) $ in terms of 
$D_{q}$ or, alternatively, $D_{q}$ if we know $\alpha $ and $f\left( \alpha
\right) .$ Some universal properties follow from these definitions. Thus $%
f\left( \alpha \right) $ is a convex function whose maximum is located at $%
\alpha \left( 0\right) $ with $f\left[ \alpha \left( 0\right) \right] =D_{0}$
and $\alpha $ varies in the interval $\left[ D_{\infty },D_{-\infty }\right]
.$ $f\left( \alpha \right) $ gives the dimension of the support of those $%
p_{i}$ that scale with $\alpha .$

As a first application of the above formalism, we consider the special case
of the transition from GOE to the Poisson for matrices of dimension $N=2.$
In this case, the probability distribution of a given component $C$ can be
worked out analytically. It has been shown in Ref. \cite{Pato0} that it is
given by

\begin{equation}
P(y)=\frac{\alpha _{0}}{2\pi \beta }\sqrt{1+\frac{\beta }{\alpha _{0}}}\frac{%
1}{\sqrt{y\left( 1-y\right) }}\frac{1}{\frac{\alpha _{0}}{4\beta }+y\left(
1-y\right) }  \label{24}
\end{equation}
where $0<y=C^{2}<1$. When $\beta =0,$ the GOE limit, the distribution is
that of the component of a two dimensional unit vector that can point evenly
in any direction. In the other limit, $\beta \rightarrow \infty ,$ the
distribution goes to a sum of delta functions and the vector is completely
localized, $y$ can then have only the values $0$ or $1$. This means that for
small values of ratio $\beta /\alpha _{0},$ the distribution is dominated by
the two power law singularities located at the extremities of the segment.
On the other hand, as $\beta $ increases, the two poles $\frac{1}{2}\left(
1\mp \sqrt{1+\frac{\alpha _{0}}{\beta }}\right) $ of the denominator,
approach the interval from the left$\left( -\right) $ and the right$\left(
+\right) $, respectively, and deform the power law behavior. In order to
perform a more detailed analysis, we partition the interval $\left[ 0,1%
\right] $ into $N_{p}$ equal subintervals of size $l=1/N_{p}.$ We have then
two kind of contributions to the partition function. Those that come from
the power law singularities at $y=0$ and $y=1$ and those from the rest of
the segment. To calculate the contributions of the first ones we integrate
from $0$ to $l$ and from $1-l$ to $1$ and for the others we just approximate
them as $\rho \left( y\right) l.$ The partition function is then given by

\begin{equation}
\chi \left( q\right) \sim \left[ \arctan \left( \sqrt{\frac{\beta l}{\alpha
_{0}}}\right) \right] ^{q}+l^{q-1}.  \label{25}
\end{equation}

For small values of the parameter $\beta $, we can assume $\beta l/\alpha
_{0}\ll 1$ and the $\arctan $ can then be replaced by its argument and the
first term becomes $l^{q/2}$. We then deduce

\begin{equation}
D_{q}=%
{1,\ \ q<2 \atopwithdelims\{. \frac{q}{2\left( q-1\right) },\ \ q>2}%
\label{28}
\end{equation}
from which we derive a $f(\alpha )-$spectrum with only two points, $\alpha
=1 $ with $f\left( \alpha \right) =1$ and $\alpha =\frac{1}{2}$ with $%
f\left( \alpha \right) =0.$ These values mean that the extremity points have
fractal dimension zero while the others have the dimension of the support.
Incidentally, we remark that we have exactly here the same distribution of
that provide by the iteration of the logistic map.

In the second situation, we assume that although the $l$ are very small $%
\beta $ is so big that we can not linearize the $\arctan $ anymore. The
first term scales then as $l^{0}$ and we deduce

\begin{equation}
D_{q}=%
{1,\ \ q<1 \atopwithdelims\{. 0,\ \ q>1}%
\label{32}
\end{equation}
and the two points $\alpha =1$ with $f\left( \alpha \right) =1$ and $\alpha
=0$ with $f\left( \alpha \right) =0$ for the $f(\alpha )-$spectrum. We
remind that we are here getting close to the situation in which the
distribution becomes a sum of delta functions concentrated in the limits of
the interval. The components approximate then the two values $1$ and $0$.
That is why the exponent $\alpha $ vanishes.

We see, in both cases discussed above, that the dimension function exhibits
discontinuity, in its first derivative in one case and in the function
itself in the other. In the thermodynamics picture of the multifractal
analysis, these discontinuity are interpreted as phase transitions\cite
{McCau} and what we learn from the above results is that the transition
chaos-order, here obtained by the variation of the parameter $\beta ,$ is
followed by a change in the qualitative behavior of the dimension function.
This modification may also be considered as a phase transition with respect
to the variation of the external parameter $\beta $. This is exhibited in
the figure where we can see that the information dimension $D_{1}$ plotted
as a function of $\ln \frac{\beta }{\alpha _{0}}$ shows a typical first
order phase transition pattern. One should expect that what we are observing
in this case of ensemble of $2\times 2$ matrices are universal features of
the transition RMT- Poisson. We pass now to the discussion of ensembles of
matrices of size arbitrarily large.

We start discussing the GOE limit. It is known that, in this case, the
probability distribution of the components is that of the components of a
unit vector in the hypersphere in the space of $N$ dimensions. It can be
proved\cite{Porter} that this is given by

\begin{equation}
P\left( y\right) =\frac{2\Gamma \left( \frac{N}{2}\right) }{\Gamma \left( 
\frac{1}{2}\right) \Gamma \left( \frac{N-1}{2}\right) }y^{-\frac{1}{2}%
}\left( 1-y\right) ^{\frac{N-3}{2}}  \label{33}
\end{equation}

The distribution is again characterized by the power singularities at the
extremities $y=0$ and $y=1.$ The partition function associated to a division
of the interval into $N_{p}$ cells of equal sizes $l=1/N_{p}$ approaches,
for small $l,$ the behavior

\begin{equation}
\chi \left( q\right) \sim l^{\frac{q}{2}}+l^{q-1}+l^{\frac{2}{N-1}}.
\label{36}
\end{equation}
and, in the limit when $l\rightarrow 0$, we find the dimension function

\begin{equation}
D_{q}=\left\{ 
\begin{array}{c}
\frac{N-1}{2}\frac{q}{q-1},\ \ q<-\frac{2}{N-3} \\ 
1,\ \ -\frac{2}{N-3}<q<2 \\ 
\frac{1}{2}\frac{q}{q-1},\ \ q>2
\end{array}
\right.  \label{38}
\end{equation}

We have therefore a double phase transition separating three states, two of
them are defined by the equation of state $\frac{q-1}{q}D_{q}=const.$ which
is generated by power law singularities\cite{Kada} and, in the middle of
them, there is the state whose equation is given by $D_{q}=const..$ For the $%
f\left( \alpha \right) $-spectrum we deduce, making use of Eqs. $\left( \ref
{18}\right) $and $\left( \ref{20}\right) ,$

\begin{equation}
f\left( \alpha \right) =\left\{ 
\begin{array}{c}
0,\ \ \alpha =\frac{1}{2}=D_{\infty } \\ 
1,\ \ q=1 \\ 
0,\ \ \alpha =\frac{N-1}{2}=D_{-\infty }
\end{array}
\right. .  \label{40}
\end{equation}
which means that the two singular contributions from the extremities of the
interval have zero fractal dimension while the rest of the others have the
dimension of the support.

The more general case, with $\beta \neq 0$ and $N>2,$ has been investigated
by numerical simulation of ensemble of matrices. The dimension function in
the limit $l\rightarrow 0$ was extrapolated from the dimensions obtained
with two small $l^{\prime }s$ by the relation

\[
D_{q}\left( 0\right) =\frac{D_{q}\left( l_{1}\right) \ln l_{1}-D_{q}\left(
l_{2}\right) \ln l_{2}}{\ln l_{1}-\ln l_{2}} 
\]
which follows from the assumption that for sufficiently small $l$, the
partition function behaves as

\[
\chi _{q}\simeq A\left( 0\right) l^{(q-1)D_{q}\left( 0\right) }. 
\]
The results obtained in this way for matrices of dimension $N=100$ are shown
in Fig. 2. We see that the structure with two phase transitions and three
states of the GOE case evolves to the picture given by Eq. (\ref{32}),
typical of the limiting distribution with two $\delta $-functions at the
extrema of the interval. The correspondent $f\left( \alpha \right) $%
-spectrum is shown in the next figure, Fig. 3, for three values of the
chaoticity parameter and matrices of dimension $N=10$. It is seen that the
singular behavior of the GOE limit, Eq. (\ref{40}) has been smoothed out by
the numerical simulation. Finally, in the subsequent figure, Fig. 4, the
first-order phase transition exhibited by the information dimension $D_{1}$,
is seen. In the same figure, it is also shown the Shannon's entropy of the
eigenstates given by Eq. (\ref{14}) with $p_{i}=\left| C_{i}^{k}\right| ^{2}$
averaged over the $k$ states. In Ref. \cite{Pato2}, the inflection point of
this entropy has been taken as a definition of the critical value of the
chaoticity parameter which separates the phase of localized and extend
states. We can conclude from this figure that this definition is consistent
with the results obtained for the information dimension.

One important point to remark here is the consequence the present analysis
has on the question of what is the probability distribution of wavefunction
components and, also, of matrix elements of an operator- strength function-
in the intermediate regime between RMT and Poisson or, in more general
terms, between chaos and order. It has been proposed that a $\chi ^{2}$%
-distribution of $\nu $ degrees freedom would fit this distribution\cite
{Alha}. However, numerical simulations seem to suggest that a combination of
two $\chi ^{2}-$distributions is necessary in order to have a good
description throw all the intermediate steps of the transition\cite{Pato3}.
The above results of the multifractal analysis seem to give a theoretical
support to this empirical observation. Indeed, what we have shown is that
the chaoticity parameter acts like an external thermodynamic variable that
induces a first-order phase transition.. Therefore, one should expect a
modification of the nature of the probability distribution as the transition
proceeds. In the $N=2$ case, we have seen that this change in the structure
of the function, Eq. (\ref{24}), is provided by the coming into action of
two poles that lie outside of the physical domain. For large size of the
matrices it is the appearance of an extra $\chi ^{2}$- distribution that
takes care of the modification of the distribution in the passage from chaos
to order.

\ {\bf Figure Captions:}

Fig. 1 Logarithm of the components of a random eigenstate for a case
intermediate getween GOE and Poisson. The calculations were done with
matrices of dimension $N=100$. A Gaussian fit is also shown.

Fig. 2 The dimension function $D_{q}$ for matrices of size $N=100$ for four
values of the parameter $\beta (\alpha _{0}=1).$

Fig. 3 The $f\left( \alpha \right) $-spectrum for matrices of size $N=10$
for four values of the parameter $\beta (\alpha _{0}=1).$

Fig. 4 Information dimension $D_{1}$ for matrices of size $N=100$, showing a
first-order phase transition as a function of the logarithm of the
chaoticity parameter. Also shown is Shannon's entropy of the eigenstates.

\end{document}